# Experiencing the Future Mundane: Configuring Design Fiction as Breaching Experiment

Andy Crabtree, Tom Lodge, Neelima Sailaja and Alan Chamberlain
School of Computer Science, University of Nottingham
Email {firstname.lastname}@nottingham.ac.uk
Corresponding author: andy.crabtree@nottingham.ac.uk

Paul Coulton
Imagination, School of Design, Lancaster University
Email. p.coulton@lancaster.ac.uk

Matthew Pilling
Lancaster Institute for the Contemporary Arts, Lancaster University
Email m.pilling@lancaster.ac.uk

Ian Forrester
BBC R&D, Media City, Salford, UK
Email ian.forrester1@bbc.co.uk

**Abstract.** This paper introduces a novel methodological approach for surfacing the acceptability and adoption challenges that confront future and emerging technologies from the perspective of mundane action, in which they will ultimately be embedded and used. This novel approach configures design fiction as a breaching experiment to surface taken for granted background expectancies that are fateful for acceptability and adoption. We explain the logic of this new interdisciplinary method and present a concrete case to demonstrate its viability: a design fiction called Experiencing the Future Mundane (EFM), which depicts a future world in which watching TV is driven by smart adaptive media. We explicate the design of the EFM, how it was configured to breach common sense knowledge and surface taken for granted background expectancies concerning how watching TV works and is expected to work, the acceptability and adoption challenges that emerge from user engagement with the experience, and how this novel approach may be adopted more broadly.

> "All designers have to grapple with the unknowability of the future. Objects that are designed here and now will come into use at some future under conditions their creator can neither know nor control … even the most mundane of acts can unravel if expected outcomes are not met." Reeves, Goulden and Dingwall [64]

## 1. Introduction

In saying that "even the most mundane of acts can unravel if expected outcomes are not met", Reeves et al. are speaking about the fragility of technological futures. The acceptability and adoption of future and emerging technologies turns on their ability to be woven into everyday situations and circumstances, into the "most mundane of acts". If they can't then they "unravel", are undone and ultimately fail as they have no traction in everyday life. We take it this is well understood in HCI. However, Reeves et al. also highlight that the acceptability and adoption of future and emerging technologies turns, *within* the most mundane of acts, on their ability to meet "expected outcomes." This reflects the common sense understanding that mundane action is expected to bring about particular results: in ordering goods online we expect them to be delivered or to be ready for collection; in using a banking app to pay an invoice we expect the transfer of funds to happen; in paying a subscription for online media we expect to be able access music or television on demand, etc. Of course, things may not always turn out as expected, but as participants in mundane action we nevertheless hold onto the prospect that everyday life will play out in particular ways and take it for granted that it will do so [32]. This means that our engagement in mundane action turns on tacit "background expectancies" [26]. That they are tacit makes it difficult to get a handle on expectations that are *fateful* for the acceptability and adoption of future and emerging technologies, as they are not usually talked about, they sit in the background, pervasive and utterly unremarkable [76]. We thus need ways or methods of *surfacing* background expectancies and making them visible if we are to apprehend the challenges that confront the acceptability and adoption of future and emerging technologies from the perspective of mundane action in which they will ultimately be situated.

Below we describe a novel methodology, which combines a social science approach to surfacing background expectancies with an arts-based design approach to understanding acceptability and adoption challenges. From a social science perspective, background expectancies may be surfaced through *breaching experiments*. These aren't experiments as the term is usually understood, but "aids to a sluggish (sociological) imagination", designed to render the "familiar, life-as-usual" character of mundane scenes and events "strange" so that background expectancies might be surfaced and detected. Breaching experiments have been used for a variety of purposes in HCI and are often marked by common interest in "making trouble" in order to surface taken for granted features of everyday life relevant to design. However, as demonstrated previously in HCI, making trouble isn't a necessary requirement of breaching experiments [19]. What is important is that breaching experiments are designed to *impugn* common sense knowledge. This might be done by depicting familiar scenes in ways that *call into question what anyone knows* about the mundane circumstances depicted and how they are expected to work, which is where an arts-based design approach enters the picture.

The depiction might be done through *world-building* [18], a distinctive approach to design fiction. Design fiction creates "materialised thought experiments that prototype near future worlds" [10]. These are not prototypes as the term is usually understood, but "diegetic" in nature [40], providing *fictional* depictions of new technologies in a future world where their existence makes sense. Design fiction has been used for a variety of purposes in HCI, often marked by a common interest in developing a "visionary story" to motivate new technological futures. However, as previously demonstrated in design research, design fiction need not be limited to narrative construction but may prototype *fictional worlds* depicting new technological objects and services through the design of hybrid assemblages that combine real and fictional technologies [18]. Both approaches to design fiction prototype future technologies as mundane, that is as practical objects used in the conduct of familiar scenes and events. The difference between design fiction as narrative construction and design fiction as world-building, is that future worlds are available to be directly experienced by potential users. The fiction is *performative*. Not in the sense of someone performing in a video to enact a visionary story [e.g., 23] or that the prototyped world consists of artefacts that play a performative role in mundane action, but in the broader sense that such artefacts are situated in an immersive scenario in an effort to "bridge the experiential gulf" between possible futures and "life as it is apprehended, felt, embedded and embodied in the present and on the ground" [14]. These "experiential futures" (ibid.) allow users to engage with the future, to interact with the fiction, and critique both the vision and the artefacts it consists of from point of view of *doing* mundane action in a posited future world.

The world-building approach to design fiction creates experiential futures that enable users to get their hands-on the future and experience it directly for themselves. However, building experiential futures that enable people to do mundane action in a posited *future* world is not sufficient to elicit background expectancies that are fateful for the acceptability and adoption of future and emerging technologies, because background expectancies are not a property of mundane action in the posited future. They are a property of mundane action *in the present*. Background expectancies are *already* built into the most mundane of acts, we expect them to play out in particular ways *now*. One further ingredient is required then if we are to elicit background expectancies: depictions of the future mundane must impugn common sense knowledge of how the mundane action depicted works and is expected to work. It is not sufficient to depict future and emerging technologies as mundane, either experientially or narratively, the depiction must also be *configured through design* to call what anyone knows about the mundane action depicted into question so that taken for granted background expectancies can be surfaced. This is not a general feature of design fictions to date.

Below we describe the distinctive sociological and design-based reasoning that shape this novel methodology and present a *demonstration* of it in action called Experiencing the Future Mundane (EFM) [60]. The EFM is an experiment in smart, AI-driven adaptive media: films or movies that adapt to the audience and are personalised based on the processing their personal data by AI, including data from smart devices, and which use smart devices to enhance audience immersion. The EFM depicted smart adaptive media as a familiar social scene: watching TV. It impugned users common sense knowledge of how watching TV works and how it is expected to work by surfacing as a key part of the experience ordinarily invisible features of a future but contiguous technological world. Findings from public deployment of the EFM show that design fiction *can* be configured as a breaching experiment and thus surface taken for granted background expectancies that elaborate acceptability and adoption challenges fateful for future and emerging technologies. In conclusion we reflect on our experience of designing the EFM to identify general strategies that combine with the socio-logic of the breaching experiment and design logic of world-building to support broader uptake of the approach.

## 2. Related work

Demonstration of our novel methodology configures design fiction as a breaching experiment to elicit taken for granted background expectancies and identify acceptability and adoption challenges that confront smart adaptive media, which seeks to put AI rather than viewers in control of personalised media experiences. The development of smart adaptive media is marked by 3 key trends: the turn to personalised content, the turn to interactive TV, and the turn to adaptive broadcast streams.

### 2.1 Personalised content

In *A Review of Smart TV: Past, Present, and Future* [2] Alam et al. note that the introduction of smart TVs brought with them the development of "content-oriented user interfaces" that require "less interactivity, browsing and searching." This was achieved through the design of "adaptive electronic programme guides (EPGs)" [48], which personalised available content to viewers. Smyth and Cotter describe one of the first *Personalized Electronic Program Guides for Digital TV* [72] and how it adapted available TV content by making *recommendations* to viewers. Recommender systems were not novel in themselves but they were new to TV, especially as a battery of new channels and digital platforms came online, presenting viewers with a large amount of choice. Adaptive EPGs were expressly designed to help users deal with the "information-overload problem" (ibid.) by making recommendations that personalise the range of content on offer and thereby filter and reduce the quantity of information that has to be dealt with. Personalisation is done by exploiting a *content-based* recommendation approach, where content is filtered based on similarity to previously consumed content, and / or a *collaborative* recommendation approach, where content is filtered based on the consumption patterns of similar viewers. Both turn on *user profiling* to build preference models of users. AI was seen as "central to the success" of personalised content (ibid.) and still is:

> "Companies such as Netflix, Spotify, Amazon Prime Video, and Disney+ now have the ability to analyse viewing patterns, genres, and user feedback to deliver personalised content recommendations and help audiences discover new and engaging titles. To optimise their personalisation strategies, streaming platforms are increasingly turning to AI app development services for expert guidance. This customisation level enhances user satisfaction, creates a sense of exclusivity and relevance, and encourages long-term loyalty. The more personalised the experience, the more likely viewers are to discover new content that resonates with their preferences. And this leads to increased consumption and higher user retention rates." [49].

Efforts to develop personalised content are ongoing and seek to support the sociality of smart TV use [3] and to integrate second screen usage into user profiling and recommendation [81]. Nonetheless, when we talk about personalised content in demonstrating our proposed methodology, it is not recommender systems that we have in mind – though an underlying reliance on user profiling is relevant – but the *content of the broadcast stream:* the program, film, movie, the thing viewed, watched, engaged with by the audience, which leads us to interactive TV.

### 2.2 Interactive TV

Interactive TV or iTV allows the viewer to interact with actual content of the broadcast stream, creating an "enhanced television" experience [48]. There are two main categories of iTV: *information interactivity*, where the viewer can access information relating to a programme (e.g., through pages of hypertext) and *participation interactivity*, where viewers can in various ways take part in the experience (e.g., vote for a contestant in a talent show) (ibid.). Participation interactivity is of relevance here. It has attracted significant attention in HCI revolving around 3 main categories: *content editing*, where the viewer plays an active role in content creation, and development focuses on authoring tools for end-users; *content sharing* between co-located and distributed users, where development focuses chat-enabled television channels and interaction / communication methods (e.g., text, real-time voice communications, VR environments and avatars); and *content control*, where development focuses on personalised recommendations, the design and use of different input and output mechanisms to select and consume content, rendering techniques, etc. [15]. Content editing is of relevance here and with it the idea that *viewers can adapt the narrative content* of a TV programme [80]. The rise of smart TV and online platforms was seen to present the opportunity to radically reimagine how content is consumed, moving beyond

"simplistic" narrative spaces where, for example, voter choices influence programme outcomes to "true interactivity" where narrative content is generated in real time in response to viewer interaction (ibid.).

> "When Netflix streamed Bandersnatch [33] – a standalone, interactive episode of Charlie Brooker's Black Mirror – back in 2018, it was hailed by some as a moon-walk moment: a small step for the streaming service, which had already experimented with choose-your-own-adventure animation for children [63], but a giant leap for television. One fan calculated that you could watch it in 90 minutes or 46.5 hours, depending on how many byways you explored on the way to its multiple endings … Five years later, the same streamer is offering Kaleidoscope, a heist series of eight colour-coded episodes that can be viewed in any order. Its boast is that there are 40,320 different routes through it." [75]

Adapting narrative content requires the development of non-linear narrative software systems that function in ways analogous, as Ursu et al. describe it, to "a personal TV postproduction team for each viewer" and thus have the ability to "instantaneously mix sources of vision and sound according to the viewer's preferences and choices" [80], which leads to adaptivity in the broadcast stream.

**2.3 Adaptivity in the broadcast stream**
Adapting narrative content requires innovation in the broadcast stream, in the ways programmes are actually put together. Initial impetus came from efforts to personalise content through adaptive hypermedia, filtering content by selectively showing and hiding content elements and links between them [12]. Mastoff and Pemberton [48] describe how the underlying logic of adaptive media could be applied to broadcast TV by treating a TV programme as consisting of multiple "streams" of information presented simultaneously to the viewer (video, audio, music, captions, etc., including multiple versions of each). Understood and treated as such, streams may be either *presented or suppressed* in different ways and at different times and for different durations to personalise broadcast content. Treating TV programmes as adaptive broadcast streams has fostered innovation across a range of areas, including efforts to reshape content presentation while maintaining narrative coherence [31], developing adaptive subtitles [29], and adaptive user interfaces that tailor content to user preferences and devices [39], exploiting viewer's personal data to drive adaptation [35], and even incorporating their personal connected devices into the experience and using viewer's interactions with them as drivers of adaptation [70], but the most crucial development, the glue that holds these and other innovations in adaptive media together, is *object-based media*.

> "The term 'object based media' refers to … … … a new class of experience in which we … encode the narrative itself: for example, a variable length or depth experience, in which the viewer (or listener) chooses how long they would like a programme to be, or how deep they would like to go into the subject matter. In this class of experience, we take a narrative view of objects, define the boundaries of objects such that each contains a useful piece of a story, define relationships between our narrative objects and connect them into a narrative graph. We can then navigate this graph to create a tailored experience that reflects an individual viewer's preferences or needs … … … our algorithms for navigating the narrative graph are built on … [a] low-level algorithm whose purpose is to respect the constraining logical relationships between narrative nodes. The algorithms built on top of this use other information within the narrative graph to tailor the user experience … looking at all the possible nodes that could be played next, sorting them based on their distance through the graph, and then selecting one based on a Gaussian probability distribution …" [4].

Object-based media and the algorithms that drive it are "close", there has been a great deal of research into and demonstrations of the approach [e.g., 55], but "commercial deployments have been slow to emerge" due to lack of commercially available tools [56] and their integration with existing production workflows [25].

**2.4 The role of AI in smart adaptive media**
While AI is and has been central to the success of adaptive media since its inception [72], its use in actually delivering adaptive media largely *remains* at the level of personalising content even today, where AI is being used to "hyper-personalise" recommendations [e.g., 21, 34]. Information interactivity is well established, as is participant interactivity but not at a "truly interactive" level where actual programmes adapt on the basis of viewer interaction [80]. There are mainstream examples – *Puss in Book, Bandersnatch, Kaleidoscope*, etc. – and lots of research demos, but widespread uptake is currently limited by the commercial adoption of object-based media. Nonetheless, the instantaneous mixing of broadcast media streams according to viewer choice can be done, it is not a research challenge. The next step is to try and *put AI in the driving seat*, adapting broadcast streams on behalf of the user, the viewer, the audience. To see if we can use AI – *rather than user choice* – to personalise programme content to viewers within the real-time flow of the broadcast. AI can already generate video [65] but it does not, as yet, possess abilities that are remotely analogous to a TV production team mixing sound and vision on the fly to deliver personalised programme content. Like object-based media, it has been anticipated for over a decade. Lee and Shin [42], for example, explored the potential of emotion recognition to exploit user's emotional state to adapt TV content, and the use of AI to drive actual content personalisation (in contrast to personalised content filtering or recommendation) is keenly anticipated by industry pundits [e.g., 1, 37, 50, 57, 59]. The idea of putting AI in the driving seat – not the user, viewer, or audience – of the adaptive media experience is novel. The possibility has recently been explored with potential users by the BBC, who commissioned IPSOS to conduct a study of audience viewpoints on the use of generative AI (Gen AI) in media [7]. The study involved 150 participants in 15 "scenario-testing" workshops across the UK, US and Australia and explored the use of Gen AI in the production and consumption of content. While seeing benefit in content filtering and producing trailers, participants were not ready to accept the use of Gen AI to create and personalise actual programme content. Our demonstration of a new methodology for understanding the acceptability and adoption challenges that confront future and emerging technologies focuses on efforts to put AI in the driving seat of adaptive media, for the simple reason that our industry partner is BBC R&D with whom we have a longstanding relationship exploring the potential of adaptive media [17, 30, 69, 70]. Our methodological approach and substantive demonstration build on this relationship and body of work.

## 3. Methodology

Here we describe the key components of our novel methodological approach. It is not confined to smart adaptive media. The configuration of design fiction, which is now commonplace in HCI, as breaching experiment may be adopted by HCI researchers to apprehend the acceptability and adoption challenges that confront future and emerging technologies more generally.

### 3.1 Breaching experiments

Breaching experiments are probably not as well known or understood in HCI as design fiction. They aren't experiments that take place in labs under controlled conditions. It would be better to think of them as social interventions. They originated in ethnomethodology, a branch of sociology that is often associated with ethnography in HCI, where they were originally used to surface the expectations that people have about everyday life and how it works [26], expectations which are ordinarily unremarkable and *unremarked upon* [76] taken for granted, known in common, and acted upon with mutual reciprocity in the conduct of everyday life rather than being discussed. Breaching experiments were devised to get people to open up, to start talking about the expectations that normally sit in the background of mundane action. Breaching experiments have been used in HCI, and not only by ethnomethodologists [19, 77], for a range of purposes including "breaching norms" of surveillance to draw the surveillance society into question [46]; to "push social boundaries" and surface potential "showstopper issues" before implementation in ubicomp application design [22]; to devise "homework assignments" that "disrupt the social norms of the home' so that we understand situated practices of computer use and maintenance [71]; to intentionally "disturb situations of everyday life" and enable reflexive insight into the social meaning of new technology [51]; to facilitate the "disruption of social rules" in design workshops and reveal opportunities for design [66]; to explore human-robot interaction in public spaces by crafting Wizard of Oz interventions that "violate what we perceive as the norm" [52]; to co-design "futuristic scenarios" that "confront existing practices with novel technology" [9]; to "create circumstances of incongruity" in virtual environments [13]; COVID-19 has even been seen and treated as breaching experiment by HCI researchers (!), where the "breach of social norms wrought by the pandemic" and "attempts of people to repair the breach through new practices and technologies" becomes a resource for design [6].

Further examples notwithstanding, insofar as the many and varied ways of using breaching experiments in HCI have anything in common it is the idea that in one way or another they *make trouble:* they breach norms, push social boundaries, disturb situations of everyday life, disrupt social rules, violate the norm, etc. This singular feature has come to define a breaching experiment no matter how it is construed or used, which is perhaps unsurprising given their original characterisation:

> "Procedurally it is my preference to start with familiar scenes and ask what can be done to make trouble. The operations that one would have to perform in order to multiply the senseless features of perceived environments; to produce and sustain bewilderment, consternation, and confusion; to produce .. anxiety, shame, guilt, and indignation; and to produce disorganised interaction should tell us something about how .. everyday activities are ordinarily and routinely produced and maintained." [26]

However, as previously demonstrated in HCI [19], making trouble and upsetting people is not a *necessary* feature of breaching experiments. Anyone who cares to look at the original breaching experiments will find that making trouble was only a feature of some breaching experiments and only for some participants, and that it is therefore an occasional not constant let alone defining characteristic. Breaching experiments don't necessarily have to disturb, disrupt or violate norms and rules, or push social boundaries either. Looking closely at how breaching experiments were conceived in the social sciences opens up new methodological opportunities for HCI, particularly when their intended *sociological function* is taken into account.

From an ethnomethodological point of view, everyday life is uniquely and distinctively a morally ordered social affair that cannot be explained in terms of rule-governed activities, including value systems, moral codes, ethical guidelines, etc. [20, 82].

"A society's members encounter and know the moral order as perceivedly normal courses of action – familiar scenes of everyday affairs, the world of daily life known in common with others and with others taken for granted. They refer to this world as the 'natural facts of life' which, for members, are through and through moral facts of life … an immense literature contains little data and few methods with which the essential features of socially recognized 'familiar scenes' may be detected … Just as commonly, one set of considerations are unexamined: the socially standardized and standardizing, 'seen but unnoticed', expected, background features of everyday scenes. The member of the society uses background expectancies as a scheme of interpretation. With their use actual appearances are for him recognizable and intelligible as the appearances-of-familiar-events. Demonstrably he is responsive to this background, while at the same time he is at a loss to tell us specifically of what the expectancies consist. When we ask him about them he has little or nothing to say. For these background expectancies to come into view one must either be a stranger to the 'life as usual' character of everyday scenes, or become estranged from them." [26]

The sociological function of breaching experiments is to affect that **estrangement**, to make the life as usual character of everyday scenes seem strange to members in order to surface how everyday life *works* and is *expected* to work by members as an enforceable moral matter.

An example to illustrate the point: The first breaching experiment. Sociology students (experimenters) were instructed to engage an acquaintance (the subject) in a conversation and insist that they *clarify* the sense of their remarks, e.g.,

"The subject was telling the experimenter, a member of the subject's car pool, about having had a flat tyre while going to work the previous day.
(Experimenter) What do you mean, you had a flat tyre?
(Subject) Appeared momentarily stunned, then she answered in a hostile way: What do you mean, 'What do you mean?' A flat tyre is a flat tyre. That is what I meant. Nothing special. What a crazy question!"

Twenty-three students reported twenty-five encounters all with similar results: Why are you asking me those questions? You know what I mean! What's the matter with you, are you sick? Why should I have to stop to analyse such a statement. Everyone understands my statements and you should be no exception! Drop dead! And so on. What strikes one immediately is that breaching experiments seem to disturb people, make trouble. However, there is rather more going on here insofar as the disturbance is provoked by the experimenters breaching the mundane ways in which conversation ordinarily *works* and is *expected* to work.

The contrast: As a preface to this breaching experiment, the students were asked to report ordinary conversations between themselves and their partners. The students found that their talk was often "specifically vague", lacking definite sense and reference, but this did not warrant the kind of clarification they sought from their interlocutors in the breaching experiment. Instead, the students found that the sense of their utterances turned on the "biography" of speaker and hearer, i.e., on the relationship that holds between them, prior conversations, the practical purposes for which they were talking and the circumstances that occasioned their conversation here and now. Their talk turned on the *retrospective* sense of meaning, coupled to the *prospective s*ense of meaning where the sense of their talk was "progressively realised and realisable" through a temporally unfolding course of "understanding work."

"The anticipation that persons will understand the occasionality of expressions, the specific vagueness of references, the retrospective-prospective sense of a present occurrence, waiting for something later in order to see what was meant before, are sanctioned properties of common discourse. They furnish a background of seen but unnoticed features of common discourse whereby actual utterances are recognised as events of common, reasonable, understandable, plain talk. Departures from such usages call forth immediate attempts to restore a right state of affairs." [26]

That's why those subject to the breaching experiment were disturbed. The insistence that they clarify the meaning of their utterances breached the mundane ways in which conversation ordinarily *works* and is *expected* to work by members (masters of a natural language who in their talk display common sense knowledge of everyday activities [28]). Non-compliance did not simply disturb those who were subject to the breach, however. As is plain to see in their responses, the retrospective-prospective sense of meaning is a sanctioned property of common discourse, when members fail to comply with it their actions are *called to account* (are you crazy, sick, everyone understands me, why not you!) in an effort to restore a right state of affairs. Now that's a very unusual thing to do because as ordinarily happens "*that* one is speaking and *how* one is speaking are specifically not matters for competent remarks" (ibid.). Rather, we get on with business of talk. However, breaching the retrospective-prospective sense of meaning throws the "enforceable character of actions" into relief: we can thus see that in calling breaches to account members seek to bring mundane action into "compliance with the expectancies of everyday life as a morality" [26]. Whether they are successful or not is another matter, but that they try do so is plain to see.

The takeaway: A series of breaching experiments demonstrated that compliance with the expectancies of everyday life as a morality is not restricted to the ways in which conversation works. Rather, and as noted above, background expectancies furnish members with a scheme of interpretation for recognising and understanding the familiar scenes of everyday life. They are, as such, "massive facts of the members' daily existence", constitutive of "What Anyone Like Us Knows", but when asked what these expectancies consist of members have "little or nothing to say" (ibid.). Breaching experiments get members to open up, to surface the background expectancies they take for granted with others, and to elaborate how familiar scenes of everyday life *work* and are *expected* to work as a morally sanctionable and enforceable matter. Despite procedural preference, making trouble is not essential to the conduct of breaching experiments. When we look back at their original description, other salient features stand out: 1) "Procedurally it is my preference to **start with familiar scenes**" – recall, as noted above, that familiar scenes constitute the natural facts of life which, for members, are through and through moral facts of life. 2) "The firmer a societal member's grasp of What Anyone Like Us Necessarily Knows, the more severe should be his disturbance when 'natural facts of life' are **impugned for him as a depiction of his real circumstances**." [26]. Start with familiar scenes impugned as depictions of real circumstances. The Oxford English Dictionary defines impugn as to "call into question." Thus, breaching experiments might be designed a) to depict familiar scenes in everyday life that b) call into question what anyone knows about the mundane circumstances depicted and how they are expected to work as morally sanctionable and enforceable matters.

**3.2 World-building**
The production of familiar scenes impugned as depiction of real circumstances might be done through world-building. World-building is a genre of design fiction. Design fiction is not new in HCI. It has previously been used to generate ideas, prototype interactions, scrutinise research agendas, and to encapsulate or communicate research findings [44]. It has its origins in art and design and is widely attributed as a concept to science fiction writer Bruce Sterling [74]. Julian Bleeker played a formative role in bringing design fiction to bear on computing, a turn he attributes in significant part to Paul Dourish [24] and the idea that science fiction often has a "shared imaginary" with science fact, which open ups the possibility of creative exploration between the two [10].

> "Science fiction can be understood as a kind of writing that, in its stories, creates prototypes of other worlds, other experiences, other contexts for life based on the creative insights of the author .. designed fictions .. can be understood similarly. They are assemblages of various sorts, part story, part material, part idea-articulating prop, part functional software. The assembled design fictions are component parts for different kinds of near future worlds. They are like artefacts brought back from those worlds in order to be examined, studied over." (ibid.)

These are not prototypes as the term is traditionally understood, they are not early versions of computing systems to be refined through continued stakeholder engagement [62]. There is a sense in the way Bleeker frames them that they are probe-like (artefacts brought back from near future worlds in order to be examined, studied over), but they are not like technology probes: while they may be part of a process of inspiring users and designers to think of new kinds of technology, there is no engineering goal to field test new technology (they are designed fictions), there is no social science goal to collect information about the use of such technology in a real world setting (they are designed fictions), and they are not deployed in situ [36]. Rather design fiction prototypes are "materialised thought experiments", "propositions done as physical instantiations" that "stand in for" and "help us imagine" what the future might look like [10].

The keyword that distinguishes design fiction prototypes from prototypes in engineering or computing or HCI is "diegetic", a notion that borrows from David Kirby's analysis of fictional depictions of future technologies in film, which he calls "diegetic prototypes" [40].

> "Diegetic prototypes differ substantially from .. 'speculative scenarios' .. such as manned missions to the centre of the Earth .. Speculative scenarios represent highly implausible and impractical situations and technologies .. imbue[d] with a sheen of plausibility, so that they look possible within a film's narrative. They make these technologies look plausible, knowing that they are impossible to achieve in real life." (ibid.)

Diegetic prototypes, on the other hand, while utterly fictional are possessed of several distinctive properties that lend them their realism. They are embodied in performative artefacts situated within familiar social scenes and are normalised through their representation as practical objects that characters interact with as if they are parts of their everyday world. They are also marked by "self-consistency in both the real world and the story world." Kirby appeals to the gestural interface in film *The Minority Report* (2002) by way of example. In addition to the above, he suggests the perceived realism of what was then a futuristic interface turns on it being consistent with the "constraints of real-world computer technologies." Computers and interfaces were already commonplace in 2002. Interacting with them by touch was not implausible or impractical, indeed it looked like "anyone could operate this technology" (ibid.). The diegetic prototype was contiguous with the real world.

Design fiction transforms the notion of a diegetic prototype from a future technology that only exists in the fictional world (e.g., on the cinematic screen or in the science fiction book), to a future technology that it is *embodied* in assemblages of various sorts – part story, part material, part idea-articulating prop, part functional software – built around the distinctive properties of diegetic prototypes to *embed* the assemblages in familiar social scenes as practical objects used in everyday life.

> "Whereas 'design' might typically highlight the object itself, outside of its .. context .. We can put the designed thing in a story and move it to the background as if it were mundane and quite ordinary — because it is, or would be. The attention is on the people .. as it should be." [10]

Coulton et al. [18] characterise this approach to design fiction as "world building" to reflect the act of placing future technology in context, making it mundane and ordinary, at home in its human world. Distinctively, world-building blends fictional (part story, part material, part idea-articulating prop) and real (part functional software) technologies to create hybrid assemblages that prototype future and emerging technologies. These assemblages do not coalesce around a story about a future world but provide an *incarnation* of it. They situate the future in, and make it visible and available as, mundane scenes in everyday life. The fiction is the hybrid world designed to be mundane.

Distinct affordances attach to the world-building approach to design fiction. Because the diegetic prototypes it makes are incarnate, embodied, users can get their hands on them and experience the future mundane for themselves [14]. The world's built and told by these diegetic prototypes are immersive and interactive, which means users can *perform* mundane action and engage in critical evaluation of the future on that basis.

> "Design fictions contain users, environments, and technologies. The technologies they contain may often be 'real' insofar as they exist in some early or prototypical form in the present, however they are 'fictional' in the way they are diegetically presented in a fabricated future .. building fictional worlds .. to describe mundane use cases in everyday situations has historically demonstrated design fiction's propensity for effectively suspending disbelief. By suspending disbelief in this way design fiction opens a potential to deeply consider the .. implications of an emerging technology." [44]

Suspending users disbelief about technological change by trying to make the future look mundane is *not the* same as understanding the expectations people have of the mundane action depicted and the *alignment* of the future with those expectations, which may of course be fateful. Furthermore, eliciting those expectations is not the same as asking people to reason about technological change once they have suspended their disbelief. Rather, we need to impugn common sense knowledge, call it into question through design, as a *built-in feature of experiencing the future mundane* enabling users to open up and elaborate whether the future works as they expect it to work and if not, to understand what problems and solutions might be surfaced in the breach. In doing so, design fiction bears something of a family resemblance to "provotyping", i.e., using prototypes as a means of "calling forth what is usually taken for granted" [53], but instead of developing and refining the usability of a functional prototype the aim is to elicit background expectancies that impact the acceptability and adoption of future and emerging technologies [44].

## 4. Demonstrating the methodology: Experiencing the Future Mundane

The core proposition – that we might usefully (and originally) configure design fiction as a breaching experiment to elicit taken for granted background expectancies that are fateful for the acceptability and adoption of future and emerging technologies – was made in a research proposal to Engineering and Physical Sciences Research Council in the UK in the autumn of 2018. The proposal, called Experiencing the Future Mundane (EFM) was successful though severely disrupted by COVID-19. Nonetheless, we managed to develop, trial and eventually deploy an adaptive media experience [60]. The EFM builds on and extends the *Living Room of The Future* [70], a design fiction that delivered an adaptive media experience personalised to the audience based on data derived from their interactions with Internet of Things devices. The EFM leverages AI and 'smart' technologies into the package and puts it on wheels – in a caravan – to provide a mobile platform that enabled the research team to take the future mundane to the public at large. Here we describe the design and deployment of the EFM and the acceptability and adoption challenges seen in and surfaced by the breach.

### 4.1 The EFM: a breaching experiment in smart adaptive media

At the heart of the EFM is a mundane scenic event: watching TV. This familiar social scene was designed into a small teardrop caravan. Despite the size of the caravan, it is possible to comfortably accommodate three adults on a sofa at the rear, with each having an uninterrupted view of a large integrated TV screen placed at the front (Fig.1)

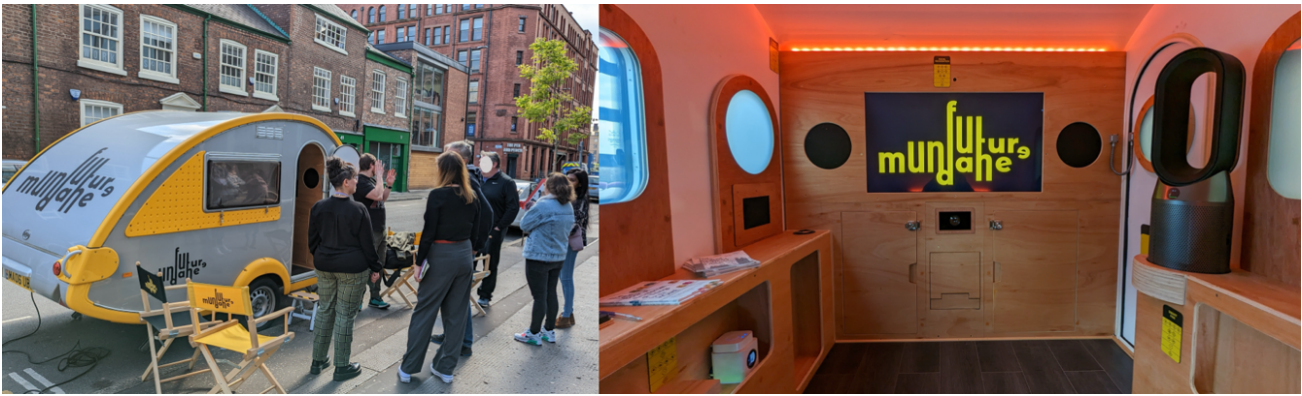

**Figure 1.** The mobile platform.

Audio speakers are also integrated into the environment, two either side of the TV screen and one below, and two plus a sub-woofer at the rear, to create an immersive soundscape enabling directional sound. IoT lighting was installed along the rear panel behind the seating, underneath the seating, and along the top of the front panelling to enable control of the colour and hue of the lighting and to match it to the changing mood of the adaptive media experience as it unfolds. These primary elements were complemented by 'smart' technologies (Fig.2), including smart glass panels in the caravan's four windows that obscure visibility and only allow low-level ambient light to enter the environment in order to increase the level of immersion within the experience; a network connected Dyson Purifier fan, which is activated to blow hot or cold air towards the audience at appropriate times, such as during a scene depicting a cold and windy location; an AI interface, actually a Wizard of Oz [38] voice-over resembling Kubrick's rendering of HAL 9000 in *2001: A Space Odyssey*, guides participants through the experience; a robotic arm with a camera mounted on it, is installed in a cupboard underneath the TV screen and appears to scan participant's faces and support facial recognition; a receipt printer installed on top of a side panel provides a printed record of the experience, including what data had been gathered from the audience during the experience and how it was used to inform the experience. A camera and microphone embedded in the front wall panel allow the research team to see and hear participants and tailor the AI experience around them. The experience was authored using an adapted version of Twine [45], which is an open-source tool for telling non-linear interactive stories [78], and in this instance allowed smart devices to be incorporated into the TV experience. The technologies used to deliver the experience run on a closed network, consisting of a computer installed behind the television and a router integrated into the left-hand side panel, to enable the caravan to operate in remote locations without internet connectivity.

This set up allows each of the connected devices to communicate and react to the live interactions and choices of the audience.

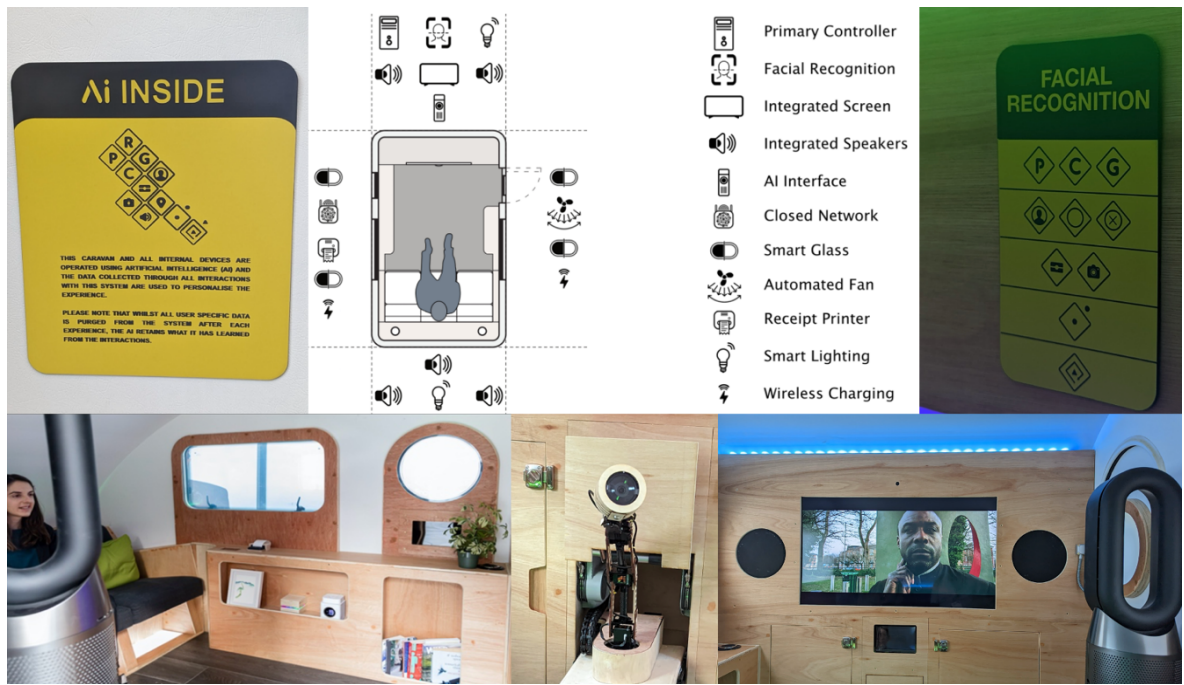

**Figure 2.** The future mundane.

The EFM is a hybrid assemblage – part story, part material, part idea-articulating prop, part functional software – an embodied fiction about a proximate technological future designed to be mundane. On entering the caravan, participants are confronted by a familiar scene: a warmly lit room, with a sofa, books on a shelf, a potted plant on top, a fan, a TV screen and built-in speakers. On closer inspection, the proximate future starts to come into view: a deck of cards on the shelf informs the user about the data the AI algorithm was trained on; a sign on the wall notifies participants of facial recognition; another that AI is inside the caravan. It states that all internal devices are operated using artificial intelligence and the data collected through participants interactions with the system are used to personalise their experience. The signage also informs participants that all user specific data is purged from the system after each experience, but the AI retains what it has learnt from their interactions with it. Riding on the tail of COVID-19, the TV screen displays live data from the Dyson fan about the air quality in the room to reassure participants that not only is the space clean (participants were told the space was wiped down after each visit), but that the air is too. The experience is introduced by a countdown on the screen (5, 4, 3, 2, 1) and then the AI speaks:

AI: Thanks for waiting. The future always takes longer than you think to arrive. I will now create a personal experience for you while protecting the privacy of your data. Data means any information that is collected about you.
Audience 1: OK

*5 second pause*
AI: This will be an immersive experience. Any questions to ask at this point?
Audience 1: No. Do you?
Audience 2: No
Audience 1: No
Audience 2: I'm good
Audience 1: OK
Audience 2: I just want to experience something before I ask any questions
Audience 1: I thought you were about to say before I nod off!
*Audience 1&2 laughing*
Audience 1: Sorry, we weren't being rude. We were just

AI: Let me tell you about
Audience 1: OK
AI: how we use data in this experience. I use sensors such as your responses to my questions to gather data about you. This helps personalise and tailor this immersive experience just for you.
Audience 1: Ooh

AI: Then I process the data I have collected about you with my artificial intelligence algorithms. These help me see patterns about the way you might behave. I will try to be as clear as I possibly can about what I am doing just so it does feel creepy or intrusive.
Audience 1: Um, that's always very nice to

AI: Do you ever worry about how the people that collect data use it, or even what they collect?
Audience 1: Yes, I do a lot
Audience 2: Yes
Audience 1: Yes, we'll go with yes
Audience 2: Yes
Audience 1: That's our business
Audience 2: Yes
*Audience 1&2 laughing*

AI: Just to be clear, all the personal data the living room generates stays in this room. It's meant to put you in the driving seat. Keep you in control. To start I need to generate personal profiles. I use this information to provide a version of a TV show that is especially adapted just for you. It will help you to be immersed.
Audience 1: OK

*3 second pause*
Audience 2: I'm ready

Audience 1: I feel like it listens but doesn't understand
Audience 2: Yeah, yeah

*Whirring noises, the cupboard underneath the TV screen opens and the robotic arm starts to emerge and point a camera at the audience*
Audience 2: Oh, oh, OK that is scary

Audience 2: I think that's a looped bit, right
*Audience 2 pointing towards TV screen where her face is being displayed and overlayed with a face scan schematic*
Audience 1: Yeah, that's looped
*More whirring noises from the robotic arm as it moves*
Audience 1: Woah
Audience 2: Oh that's weird. Where are you (to audience 1)?
*Audience 2 waves at TV screen*
Audience 2: OK, my privacy has been compromised

AI: I have been watching you do this. I'm part of the TV
Audience 1: I've been watching you too
AI: I've been scanning your face since you arrived and very nice it is too. Don't worry about it because data never leaves this space.
Audience 2: I'm feeling extremely self-conscious right now
*More whirring noises as robotic arm turns to audience 1*
Audience 2: Oh
Audience 1: Just you, not me
*Robotic arm making more noises as it moves*
Audience 1. Just you, not me!
*Robotic arm retracts into cupboard*
Audience 1: Bye bye

AI: Oh wait. One minute. Facial analysis shows you dislike my voice. What about this? (voice changes from male to female). Or this (changes to a different male voice). Oh no, you really don't like that, what about this (female voice again)? Don't you like me then (back to different male voice).
Audience 1: Is that Australian?
Audience 2: Yeah
AI: I'll go back to being myself (original male voice)
Audience 1: OK
AI: Much better
Audience 2: OK, alright
Audience 1: Yep

*3 second pause*
AI: Based on your generally agreeable features
*Audience 2 laughs*
AI: I now conclude that you consent to further profiling of your facial features
Audience 2: No, I was worried
AI: Yes?
Audience 1: Yes
AI: OK. Initial analysis of your facial features suggests that you're a sucker for a romantic story. You're in for a treat The drama will be right up your street
*Audience 1 looks at audience 2; audience 2 laughs*
Audience 1: I am a sucker for a romantic story, how did it know!

AI: I have bad news, as it happens. I have used an anonymous face profile and compared it against a health service database. There is a medium to high risk that you will get heart disease in the next ten years
*Audience 2 laughs*
AI: I recommend taking a walk after this experience
Audience 1: Probably that pizza
Audience 2:           that pizza!
*Audience 1&2 laugh*

AI: Right, now I am just going to explain the sensors in the room. This is necessary for our meaningful consent procedures. Alright?
Audience 1: Yes
Audience 2: Alright
AI: If you wish to withdraw your consent, I suggest you hold your thumbs up now
Audience 1: No
Audience 2: To withdraw
Audience 2: No
Audience 1: did it say?
Audience 2: Yes
Audience 1: OK, no

AI: So, back to the sensors. I run all the data through an artificial intelligence algorithm. This allows me to provide a complete personalised media experience. Assuming that all makes sense, then if you are happy to go ahead say yes.
Audience 1: Yes
*Loud noise*
Audience 1: Is that in, outside?
*Articulated vehicle passes the caravan outside*
Audience 2: Yes
*The windows turn from opaque to clear*
Audience 1: Woah
Audience 2: Oh what, oh my god
*The windows turn opaque again*
AI: I can block out the light with my smart windows
Audience 2: Oh wow
Audience 1: Wow
Audience 2: I like that
Audience 1: Yeah
*Dyson fan turns and starts moving*
AI: I can give you a nice blast of air
Audience 1: Beautiful
Audience 2: Yeah, I think she really needs that
Audience 1: I do
Audience 2: After the
*Lights in the room start to change colours*
AI: I can even make the room all moody
Audience 1: Ooh
Audience 2: OK
Audience 1: Very moody
Audience 2: Yeah
AI: And of course I can show stuff on the good old TV screen
Audience 2: OK, we're ready

*Receipt printer makes a noise and prints a line on the receipt; audience 1&2 lean forwards to look at it*
AI: Please tear off your receipt now and sign it on the dotted line with the pen provided
Audience 1: Tears off receipt
Audience 2: I think you should sign (laughing)
Audience 1: Me! It's your face
Audience 2: I don't want to sign for that heart disease thing
Audience 1: Come on, there you go
*Audience 1 passes receipt and pen to audience 2; audience 2 signs the receipt*
Audience 1: Ah
AI: Brilliant. Now, what we have all been waiting for. Based on you and your data I am going to show you a personalised version of a new drama series called A Few Moments. Sit back relax and enjoy the immersive experience I have created just for you
*Ambient sounds come out of speakers, lights start changing colour, fan starts blowing air, audience 1&2 look at the TV screen as the drama begins*

This sequence of interaction shows the reader what a typical introduction to the smart adaptive media experience looks like. It's a composite, seven minute sequence drawn from three different audiences to give the reader a flavour of the kind of ordinary interaction that routinely took place between audience members and the AI in the caravan. After the drama started, audience members typically engaged in "TV talk" [41] as part of their mundane interactional immersion in the drama: mundane conversation of the kind you might find in any home where people are sat watching TV together; e.g., "What's this about?" "I feel like I recognise him." "From where?" "I thought that guy was waiting" "Oh, they've found each other", etc. As the drama unfolds, various connected objects in the room contribute to the audience's immersion in the experience and the sense that the experience is being driven by AI. The room's lighting adapts to a colour gradient matching each scene, for example. When the lead character in the short drama is outdoors, the fan switches on in response to the wind blowing her hair and blows the participants' hair. Captions at the bottom of the screen indicate when data is being collected in the room and used in the experience. Lighting reinforces this, visually displaying that connected devices are transmitting data. The music within the drama, and the ending of the drama itself, are selected on the basis of the putative profile. At the end of the drama, the receipt printer quite literally spells out how it was uniquely tailored to the participants through the collection, analysis and use of their data to drive the experience.

*None of this actually happened.* Of course participants watched TV, the lights changed around them, the fan blew, the windows went opaque, etc. But none of it, and none of what went before it, was actually driven by AI. It was Wizard of Oz [38], mocked up by humans to make it look like an autonomous system was at work. A designed *fiction*. There was no facial recognition, no profiling. Where technological effects were experienced, then they occurred because they were woven into Twine and chosen on an ad hoc basis by the operators situated outside the caravan (Fig.3) watching, listening and responding to what goes on within to give the impression that the AI was tailoring the experience uniquely for participants. Nonetheless, the fiction wasn't perceived as such. Some found the experience a bit confusing, a bit random, rudimentary, mocked up, staged, a kind of theatre where participants were unsure sure how much was scripted as the experience didn't seem all that personalised, but rather that someone was behind the AI, more like people talking through machines. Others found the experience interesting, engaging, really quite cool, playful, amusing, humorous, fun, and the AI more human than they expected or had experienced to date. Either way, no one thought the experience a mere work of fiction. On the contrary, as the findings from our post-hoc interviews made clear, whether entertaining reservations about the experience or embracing it, the EFM was very much experienced as real.

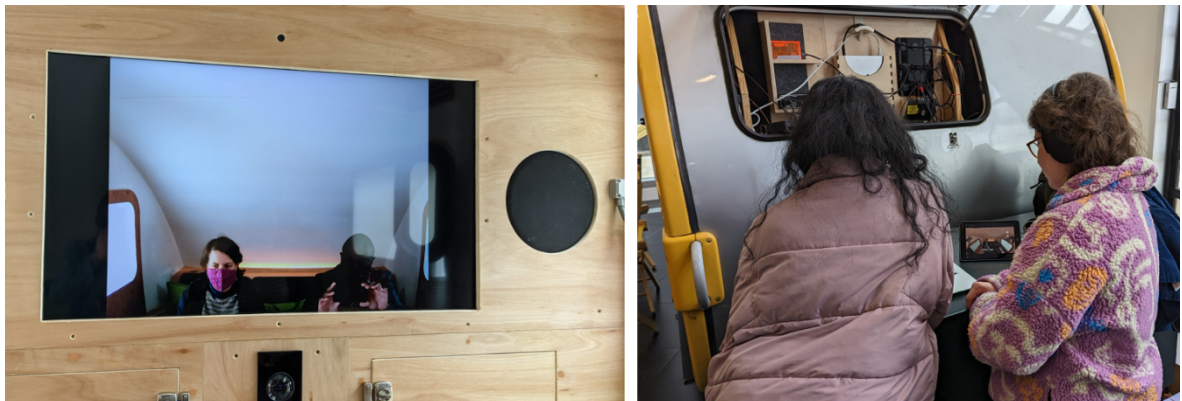

**Figure 3.** Outside looking in: members of the research team operate the experience.

But what about impugning common sense knowledge and eliciting background expectancies? As can be seen in the sequence of interaction above, participants weren't always comfortable with the experience: that's scary, weird, I'm feeling extremely self-conscious right now, just (do it) to you not me! We weren't trying to disturb, disrupt, or violate social norms, etc. We were trying to call what anyone knows about watching TV and how it is expected to work into question. We tried to do this by *surfacing through design* ordinarily invisible features of future technological infrastructure: the placement of signage to surface that AI was inside the room; the AI voice surfacing personal data processing and profiling; the robotic arm, camera and TV screen surfacing facial recognition; the use of lighting to surface data transfer from connected devices; the on-screen captions surfacing the ongoing and live nature of personal data processing; and the receipt printer surfacing what data was used to personalise the experience. Why did we think these features of the technological infrastructure will be invisible in the future? Because they are now: no one tells us AI is the room when we enter someone's home, and data processing and profiling is limited to privacy notices and consent of the account holder. Contiguity underpins the fiction and it was in designing the future mundane so as to surface the ordinarily invisible nature of technological infrastructure, both literally invisible insofar as it is embedded in other things (e.g., connected devices) [73] and / or practically invisible insofar as it disappears in use [77], that we sought to impugn common sense understandings of how watching TV *works* in the proximate future and thus configured the design fiction as a breaching experiment in the here and now.

**4.2 Deploying the EFM**
Following long interruptions to development occasioned by COVID-19, trial runs to see if the technology actually worked [60], and then finding venues to host it, the EFM was eventually made available to members of the public at two events in 2023. One in May at the University of Salford's Media City building; Media City is a major centre of TV production in the UK, located at Salford Quays in the North of England. The other in June at the Bath Digital Festival, a community event in the south-west of the country that raises awareness of new technologies and provides a platform for community members to build connections locally and nationally. In both cases, the caravan was positioned close to a café to attract passers-by. A broad range of participants was subsequently recruited from the flow of people into and out of the cafés, from those who work in media production through to technology developers, students, university staff, event attendees and members of the public at large who were having a day out at Media City or the Digital Festival. We do not categorise our participants further (e.g., in terms of demographic information or participant familiarity with AI) as we are not conducting an experiment in the classical sense of the word and have no need to manipulate variables. Rather, in conducting a breaching experiment and as described in detail above, we seek to impugn common sense reasoning through the design and subsequent engagement of participants with the EFM in order to surface What Anyone Like Us Knows about the mundane business of watching TV and how members expect it to work here and now and in the future. The imposition of categories extraneous or incidental to the mundane business of watching TV would not help us surface the background expectancies that members hold and the acceptability and adoption challenges that accompany them. The Wizard of Oz nature of the experience was not disclosed to participants at either location during or after the experience. Post-experience interviews were conducted with 30 participants over 3 days at Media City, and 20 participants over 2 days at Bath (50 in total). Participants could experience the adaptive media show without taking part in the research. Those that did agree to take part were taken through our ethics process before entering the caravan and were interviewed on leaving the caravan. The interviews were conducted on the same basis as participation, so individual participants were interviewed individually, couples in pairs, parties of three in a group. The interviews were semi-structured in nature and probed participants reflections on the experience, their reasoning about putting smart adaptive media in their own homes, and their reasoning about the need or not to control smart adaptive media. The interviews lasted around 15 minutes each and were recorded on audio for subsequent transcription and analysis. This might be thought too short a time to garner sufficient insight, though the findings below demonstrate otherwise. It might also be asked why we didn't analyse video of interaction in the caravan instead or as well? While it is certainly the case that video captured moments when common sense reasoning was impugned, it was within the ongoing and unfolding flow of an experience that provided no opportunity for participants to *discuss* the background expectancies that were being breached. Post-experience interviews provided opportunity for participants to reflect and explicate those expectations.

Our analysis of the interviews attends to endogenous topics of mundane reasoning [61], i.e., topics participants themselves invoked to describe the experience and respond to our questions, and leverages the documentary method of interpretation [27] to identify patterns in the data. The documentary method of interpretation sits alongside breaching experiments in Garfinkel's seminal text *Studies in Ethnomethodology* (ibid.). It is apposite that we use it here, though other analytic methods such as thematic analysis [11] commonly employed in HCI could be used instead. The documentary method resembles thematic analysis to some extent, insofar as it entails considerable interpretive work "at the intersection of the researcher's theoretical assumptions, their analytic resources and skill, and the data themselves" (ibid.). However, the documentary method does not code data in order to develop themes. Instead,

> "The method consists of treating an actual appearance [e.g., an utterance or sequence of talk] as 'the document of', as 'pointing to', as 'standing on behalf of' a presupposed underlying pattern. Not only is the underlying pattern derived from its individual documentary evidences ... the individual documentary evidences, in their turn, are interpreted on the basis of 'what is known' about the underlying pattern. Each is used to elaborate the other [27]."

The documentary method of interpretation is an ordinary analytic method that turns on the mastery of natural language [28] and with it the *hearer's maxim* [68], which holds that conversational topics that are hearably consistent with one another "go together" (ibid.). In going together, individual documentary evidences (topical utterances in transcripts) begin to elaborate an underlying pattern, and the pattern in turn is used to interpret individual documentary evidences. So in contrast to coding, using the documentary method of interpretation to analyse interview transcripts entails *clustering* comments that hearably go together to identify patterns where distinctive topics of mundane reasoning coalesce. Our findings do not present, define and elaborate particular themes then. Rather, they blend together words, phrases, utterances, and sentences from the data that hearably coalesce to elaborate distinctive topics of mundane reasoning surfacing the background expectancies participants hold about watching TV, here and now and in the proximate future. The findings presented below are not simply to do with smart adaptive media. They provide an empirical *demonstration* of the *efficacy* of the proposed methodology. They show that design fiction *can* be configured as a breaching experiment and surface taken for granted background expectancies that elaborate acceptability and adoption challenges fateful for future and emerging technologies.

**4.3 What's seen in the breach**

It came as no surprise that our participants were already familiar with smart technologies. Everyone had a smartphone, many had smart speakers, a few had smart TVs and smart thermostats, smart lights, the odd smart watch, one person had a robot hoover, another had networked attached storage for serving media and also "programmed things with Raspberry Pi's." This background familiarity with an everyday world inhabited by an increasing array of smart things furnished a general scheme of interpretation that framed participants engagement with the EFM.

> P7(2)M: We're cyborgs already. We're augmented humans, aren't we? How I find my way about, how I plan my day, how I do lots of mundane tasks, even pre-AI, are heavily standing on the benefits of technology.

The existence of the augmented human was commonly seen and understood to rely on widespread data gathering. Our participants told us our data is being collected a million different ways already. Everything we do at this point is tracked. Whatever we speak is being overheard by Google. You start the Internet and they have access to all my data. They've got my heartrate, my biometrics every second of the day. There is nothing that is not shared today. They already have access to our data totally. That's not an illusion, that's the truth. A truth already grounded in AI: news, as well as the songs, wherever we shop, whatever we select, the offers whatever, it's already AI controlled. This was seen as an inevitable condition of everyday life, impossible to avoid. Participants thought that while people still care about their data, there is nothing they can do and so they have given up the fight and try to forget about it, instead accepting widespread data gathering and AI as necessary to daily participation in and engagement with society, "part of the covenant we have now." Participants were not surprised by the idea of a proximate future in which personal data and AI is used to drive media experiences, then, as it's a taken for granted background expectancy that this is how everyday life now works, part and parcel of the new social covenant.

Nonetheless, our participants expressed some discomfort about the EFM when asked to reflect on the experience.

> What did you think about the experience of the caravan?
> P17B: I think it's cool that you have a custom designed living room and all the machines can design or make you feel more comfortable, but it's not on an emotional level, and that's what I don't like. Because it's like, what am I doing? I'm in front of a machine telling me what I like and what I want to be. I can't relate to that.

Participants said that the smart adaptive media experience made them feel weird and odd because the AI was in control, despite it insisting that they were in control and that its actions were merely being driven by their behaviour. While participants liked the experience of being in a futuristic living room that adapted to them, they felt they couldn't trust that it wasn't connected to a large corporation that sought to exploit them in some way. Participants were a little bit afraid of a future in which machines not only talk to them but also analyse their actions. "What did I do? What did they see on my face to react like this?" The EFM was unnerving, scary, and did nothing to alleviate fears of AI. It disrupted participants' sense of reality. "I don't want to be part of this world. How will we ever know what is real in the future?" The central idea of using smart technologies and AI to personalise media experiences was also found to disrupt the social fabric. "That's not a positive. I'm interested in watching the same show and discussing that with someone who has a completely different experience."

Our participants discomfort became more pronounced when we asked if they would like to have smart adaptive media experiences in their homes.

> P30M: Well, I can see it being quite useful. I mean I quite like the concept of Philips Hue, for example. I like the fact that it sort of adds immersion to watching films or playing computer games or whatever it might be. I like that concept. [It's] the element of data gathering that makes me a bit anxious.

Anxiety palpably turned on who was gathering the data and who the data was being sold on to, because as anyone knows that's how the model works. This model was found to provide limited opportunities for consent, particularly controls on third party data processing are not good enough. Participants also acknowledged, as a feature of the current model, that they were routinely being profiled as part of their ongoing commodification. They were uncomfortable with the idea of smart adaptive media profiling them any further so that they might receive personalised experiences from third party providers. "I'd be happy for them to share that data with devices which tailored my experience, but not other third parties or services." Extending the premise that they are always being listened to by their smart devices, participants also felt uncomfortable by the idea of being watched constantly by AI. "If they understand what I say, if they know how I feel, like see it on my face, I wouldn't like to have that they can know how I feel during the day." The use of data derived from household devices was also seen as potentially dangerous, impacting the content participants would be offered and fundamentally affect how they see the world. "Then what do we trust? Where do we know where the truth lies? That's the thing that worries me." Adaptive media was seen to undermine choice and human creativity too. "I don't want to farm out my human experience in that way. I would prefer to make those choices for myself. I fear the degradation of our own free will." The potential impact on choice and free will also extends to the ability to relate to others and share one's experience, which is fatefully undermined "if you have something that's been designed specifically for you." Additionally, participants were concerned about the potential economic impact of smart adaptive media. "Any new technology you bring into your home, the more new it is the more at risk it is of failure and then it's expensive. So it could economically be quite impactful if the equipment fails." Thus, while accepting the new social covenant as condition of participation in society, the vast majority of participants were not inclined to adopt smart adaptive media in their own homes. "I feel like there is a limit, like the use of personal data for personalising entertainment is probably a step too far for me at the moment", "I think it's yet to prove itself to be reliable and trustworthy."

Trust in smart adaptive media turned for our participants on two on two key factors, the first being the market. Views on the market encapsulated a breadth of viewpoints from the downright distrustful ("Company A can be bought out by company B and then where's your data gone?"), to the sceptical ("Is the reason they're gathering data for their benefit to make money off you, or is it specifically to improve your experience?"), to the sanguine ("If I knew that data was going to companies that I had previously bought from and trusted, I'd be like, oh, it's fine."). Of course that's a big IF and while participants were willing to trust companies perceived to have good intentions and the transaction of data clearly adds value to their everyday lives, they were for the most part not convinced that the market has their best interests at heart. There is, then, something of an existential tension between the augmented human and a market that would serve them smart adaptive media. On the one hand, the augmented human must share data to participate in contemporary society, but on the other, the augmented human often lacks trust in the parties their data is shared with. How might the tension be resolved?

> How would you feel about having this sort of stuff in your home?
> P5M: I quite like that, as long as the AI was mine and not the company's, does that make sense? Like as long as the AI was like under my supervision, my control, as long as it's under my control all the time.

Control was seen to be essential, given the perceived threat posed by the market and AI to human agency and autonomy. Participants didn't want to be "diminished to a piece of data", mere homunculi in the machine. They didn't want their ability to explore other stuff to be limited by AI. They didn't want to be fed TV shows that were entirely driven by AI, but to retain "some sort of personal curation" in the media they were exposed to and consumed. This was and is a moral as well as practical imperative driven, as touched on above, by the potentially dangerous impact of smart adaptive media on participants' perception of reality, coupled to the AI's perceived inability to tell right from wrong. For example, "and to be kind of provocative, you don't want the pornography you watch to affect the recommendations you get on iPlayer, you don't want those worlds to interact." Rather, adaptive media needs to find the right balance between the use of AI to improve the viewing experience and the choices human beings make and wish to keep making.

Finding the balance turned for our participants on reimagining the role of AI. Instead of automatically making choices based on their actions, it would be better to let AI assist in their decision-making. "If they were able to offer it as almost like a concierge service, the better it knows me, the more it can service my needs." In return, participants wanted more visibility of the data they are giving up. They wanted it to be made really overt what data is being recorded, what it's being used for and when it's going out of the home at all times so they could make informed decisions. Transparency was not limited to data gathering either, but also extended to analytics, to what AI does on the basis of the data it has ingested.

> P19B: Often we don't exactly understand how something is making a decision. If I had more of an understanding and it was more transparent about where these things have come from, then I would start to feel more comfortable.
> P4M: The ability to understand what conclusion an algorithm has drawn is really important. Like I really loved that receipt that sort of goes into some of that, like it analysed your mood, I'd like more detail in that respect.

Greater transparency is necessary needed, then. Privacy notices that prospectively account for the use of data are not sufficient. Live real time views on data processing and retrospective accounts of automated decision-making are needed too.

While necessary, transparency is not a sufficient control in itself either. Direct control over data processing by AI is also required.

> Would you like to control what AI does with your data?
> P22M: Absolutely, yes.
> What would your priorities be there?
> P22M: To be given the control in the first place.

Acknowledging the new social covenant, participants nevertheless wanted to be "back in control" of their data where smart adaptive media is concerned and entertained a spectrum of views on what this entails. Prospective control mechanisms should allow participants to turn data processing off entirely, a kill switch that would terminate AI's connection to any of the participant's data and sources of data. Controls should allow participants to select data sources that can be shared and those that cannot, starting from a zero base level (i.e., nothing shared by default). "I want an opt out completely option and with individual opt ins, rather than when they're all ticked and you have untick every single one." Controls should allow participants to specify presets where particular kinds of data source are used to deliver particular kinds of adaptive experience (e.g., the use of Hue lights to deliver ambient lighting). "I would willingly give up continual control at that point because I'd preset what I like." Controls should also allow participants to determine what decisions can be made by AI and what cannot, "and would it have to get my approval before implementing any decisions." Controls should provide a database of recipients allowing participants to see who has access to data sources and why. Controls should allow participants to manage where data is going and to revoke access and delete data. "You say, now forget everything about me, forget you ever knew me – I would make use of that and it would encourage me to share more data because I knew it was just, you know, once that transaction is done, it's like it never happened, right?" Controls should also extend beyond the individual to trusted groups who assess the trustworthiness of smart adaptive media services on behalf of the individual. "There needs to be human oversight somewhere in that because otherwise, you know, we're taking away our own agency."

**4.4 Surfacing acceptability and adoption challenges**

The point and purpose of designing and deploying the EFM to breach mundane expectations was to understand the acceptability and adoption challenges that confront smart adaptive media. It is towards surfacing these challenges, as elaborated by our participants in their reflections on experiencing the future mundane, that we turn here. As can be seen in participants responses, our experiment successfully impugned (called into question) common sense understandings and expectations of how watching TV works as a mundane, morally sanctioned accomplishment in everyday life. It is plain to see that while participants expect personal data to be gathered and used as part of increasingly smart, AI-driven service provision in general, as this is a default feature of the new social covenant, the EFM nevertheless breached common sense reasoning about how watching TV works and should work in the proximate future. Thus, participants did not expect intelligent machines to subject them to surveillance in their homes and use that data to choose then personalise what they watch on TV. That would be odd, weird, unnerving, scary, and dangerous. It would have fundamental effect on the perception of reality, degrade free will, and disrupt the social fabric. Participants did expect their ongoing commodification if such technology was allowed into their homes and had no desire to let an untrustworthy market have more of their data and use it to control what they watch on TV. In short, what can be seen in breaching common sense understandings of how watching TV works and is expected work in the proximate future is that AI-driven media and the current market model are significant barriers to the acceptability and adoption of smart adaptive media. We can see, too, that there is hope. Some possibility that smart adaptive media experiences could become part and parcel of everyday life *if* the vision of the proximate future is *reconfigured* in line with expectations of how it should work in the future. Based on the participants comments in the previous section we can thus say that acceptability and adoption turn on the following:

- *AI as human assistant (not driver).* It is clear that many participants found adaptive media engaging, interesting fun and enjoyable. The challenge to acceptability and adoption is AI being in the driving seat. Enabling human control over adaptive media experiences – including the devices and data used and the decisions AI can make – is key to addressing the challenge. AI should assist humans in delivering smart adaptive media experiences. Learn from human behaviour within the constraints that humans set.

- *Enhanced transparency.* Greater transparency is required if people are to trust smart adaptive media experiences and adopt them in the home. The current market model of providing privacy notices that prospectively account for data gathering, processing and third party sharing and which require consent prior to use should be

- *User controls*. Acceptability and adoption also turn on furnishing users with a spectrum of control mechanisms describe above that provide direct control over the data sources (including smart devices) used in adaptive media experiences, data processing and sharing, and data analytics (automated decision-making). Controls should be default off (no data sources selected), enable fine-grained selection, revocation of access, and include a kill switch. Controls should also enable users and their data to be forgotten by a smart adaptive media service.

- *Human oversight*. A final challenge to acceptability and adoption lies in the provision of human oversight. What this means is less clear, but involves the evaluation and approval of smart adaptive media services by trusted parties. Whether that would be trusted parties within the marketplace (e.g., public service providers) or without (e.g., media regulators) is uncertain, though of course there may be space for both. It is also the case that oversight will be impacted by regulation (e.g., the EU's AI Act [5]), as the fundamental ability of smart adaptive media to influence human perception of reality is clearly high-risk and poses a threat to fundamental rights [79].

Acceptability and adoption challenges surfaced by our breaching experiment raise social and technical requirements for data protection and the use AI to deliver smart adaptive media experiences. While participants in our experiment may accept a new social covenant that trades on the sharing of personal data, it clearly has limits and there is evident need to think carefully about the role of AI, the provision of controls over data processing and the operation of AI, and how oversight of smart adaptive media experiences might be provided.

## 5. Discussion

As shown in considering related work in section two, smart adaptive media where AI rather than user choice drives real-time adaption is novel, a body of research doesn't yet exit with which to compare our findings. Nonetheless, there are parallels and connections with prior work, and readers may well draw their own too. Our findings clearly touch on work related to corporate surveillance in the smart home [16], the impact of content filtering on persons perception of reality [58], trust in AI [8], the role of AI assistants [47], privacy, transparency and control in personal data processing [54], and human oversight of AI [79]. These are some of the most significant socio-technical issues of recent times and they span multiple application domains. We do not offer global panaceas here, but our breaching experiment does provide domain-specific insight that goes beyond existing findings of the BBC / IPSOS study discussed in section two [7]. The study highlights general concerns with the impact of generative AI (Gen AI) on representations of reality, the use of Gen AI as an assistant, transparency in data processing, and human oversight. With respect to anything that might resemble smart adaptive TV media, the study found that participants were uncomfortable with the use of Gen AI to contribute to part of an overall piece of content and saw the use of Gen AI to create the entirety of a piece as high-risk. However, this relates to media production not the personalisation of broadcast content to the audience, which was seen as "much more challenging" (ibid.). Where personalisation is concerned, participants were only comfortable with Gen AI being used to create personalised recommendations and video trailers or summary clips. Participants wanted the role of Gen AI in creating storylines to be limited, feeling that it was a risk to human creativity and connection. Participants were comfortable with Gen AI being used as an assistive tool to support human creation, but not autonomously making, producing and directing a TV show. The need for human oversight of content made by AI was seen as essential.

There are similarities in these findings, some resonance with our own, but they are not the same and not as rich, the reflections offered by participants not as deep or far reaching. It was observed in review that our participants are "well-articulated regarding technology in their reflections." This is not due to the selection process, but a direct result of our novel methodological approach. People's expectations of everyday life and how it works are, as we explained in introducing breaching experiments, *unspoken* in the ordinary course of events. We do not ordinarily go about our everyday lives discussing the new social covenant: the wholesale harvesting of personal data, its use by companies, the trade-offs and problems involved, etc., as a condition of participation in contemporary society. Rather, we get on with the business of taking part everyday life in much the same ways everyone else. Talking about the role or impact of technology within everyday life has to be occasioned. The EFM provided one (breakdowns, new purchases, children's homework, topical interviews, etc., provide others). The EFM was an occasion that impugned and was *designed* to impugn participants expectations of how the familiar everyday scene of watching TV works. Given the technological nature of the designed experience, coupled to widespread pre-existing reliance on digital technology to consume TV media, it is little surprise participants had much to say about technology in their reflections and not only about the vision of smart adaptive media represented by the EFM. In short, the breaching experiment *worked*, it was effective, surfacing ordinarily unremarkable, unsaid expectations about how watching TV works in the here and now *and* how it is expected to work in the proximate future. Hence a wealth of "well-articulated" reflections surfacing a range of acceptability and adoption challenges only partially touched on by the BBC /IPSOS study, partly because its scope was broader and partly because it was done through the use of traditional scenario-based methods commonplace in HCI [67]. Thus, configuring design fiction as breaching experiment opens up new methodological opportunities for HCI that extend beyond the substantive case provided here.

As any methodology, ours has its limitations. It is naturally suited to eliciting acceptability and adoption challenges confronting future and emerging technologies intended to be embedded in familiar scenes of everyday life. Commonplace scenes, "massive facts of the members' daily existence", constitutive of "What Anyone Like Us Knows" [26], like watching TV. There is an in-built generality to such scenes. The expectations members have of them are "socially standardized and standardizing", used as a common sense "scheme of interpretation" (ibid.). Breaching them thus surfaces what for members are generalisable features of familiar social scenes, which impact the acceptability and adoption of future technologies within those scenes. However, this is not to say that the findings generalize *everywhere*. Where familiar, life as usual events are different – in places where watching TV is a different experience, for example, e.g., much more communal and not necessarily based in the home – then we should expect the background expectancies that people hold and use as a scheme of interpretation will be different too. We might speculate, then, that taking the EFM to rural India or Africa, for example, will produce very different results. Indeed the experience may even breakdown given the lack of *contiguity* between our world and theirs. We would have to design a different depiction to impugn members understanding of this familiar everyday scene if wanted to elicit background expectancies in countries very different to our own.

How we configure design fictions as breaching experiments is of course an important question. We do not offer guidance on how to create or conduct design fictions, including our favoured 'world-building' approach, as a substantial literature is already available. Rather, we reflect on our demonstration to identify general insights. We set about breaching mundane expectations of how watching TV works by *operationalising a concept* of a novel technology we wanted to explore with ordinary people, people whose lives we envisioned the technology being embedded in, and whose viewpoints on the acceptability and potential adoption of future technology we sought to elicit. Operationalising the concept of smart adaptive media entailed *crafting a depiction of a familiar scene* that enabled participants to get their hands on and actually experience the envisioned future. What is crafted is not simply watching TV in the future for example, as design fiction might ordinarily do, but crafting watching TV in the future to look like a familiar scene *here and now*. The crafting turns on an *interdisciplinary or multidisciplinary mix of viewpoints*: e.g., on our industry partner's *vision of the future*; on computer science and understanding what was technologically possible and, equally important, what was not, and thus on *understanding what can be built into the experience and what will have to be created fictionally*; on HCI and *configuring the interaction* between the envisioned technological future and the human actors involved; on ethnomethodology and *understanding the mundane business at hand* (e.g., watching TV)**,** what it looks like and ordinarily consists of for members here and now; on world-building and *designing the future to be mundane*. Crucial too was the *intention to breach* mundane expectations about the ordinary, commonplace business at hand. This reached beyond disciplinary contributions and required us as a team to explore ways in which we could design the experience to impugn (call into question) common sense understandings of how the business at hand works and is expected to work as a familiar scene in everyday life. We arrived at the idea of impugning common sense knowledge in the case of smart adaptive media by surfacing the ordinarily invisible features of the underlying technological infrastructure, real and fictional: processing personal data to profile users is already commonplace, for example, content filtering turns on it but it sits in the background of media experiences, just as we anticipate the future use of data from personal devices, facial recognition, AI and automated decision-making will. We could have designed the experience to hide the technology, but bringing it to the foreground and making it an explicit part of the experience allowed us to create both a sense of contiguity (it still looks like watching TV) *and* to create tension between the here and now and the proximate future (but not as we know it). Had we left the technology in the background, there would have been nothing to engage participants accept the AI voice, which would have resulted in a thin experience, not rich enough to call into question what anyone knows about how watching TV works and is expected to work.

The question was asked in review, "Could there be other ways?" Yes, we imagine there may be many, and a caravan isn't required to configure design fiction as a breaching experiment either; the caravan was used to reach out to and engage the public, not to breach their expectations about watching TV. It is the *designed experience* within the caravan that breaches background expectancies and surfaces acceptability and adoption challenges, and it is the designed experience that matters generally. While we have no cookbook recipe to offer, as configuring the designed experience to breach expectations will always be *dependent* on the future or emerging technology under consideration and the familiar scenes in which it is envisioned to be embedded, we can suggest **four general strategic issues** for researchers to consider based on our experience to date:

- **Conceptual conversation.** The format of the breach – i.e., the particular form depictions of familiar technological futures take – should be shaped with respect to the technological concept under investigation *and* the conversation the research team wants to engage participants in with respect to that concept. The design format of the EFM reflects the core idea that AI, not the viewer, adapts media based on the collection and ongoing real time processing of personal data derived from the audience and their environment, for example. We wanted to explore the acceptability of this concept with ordinary people and *the design format was (and should be) configured to enable the conversation*.

- **Collaboration.** Design concepts don't materialise fully formed. Our industry partner's idea of smart adaptive media doesn't elaborate the EFM design format, for example. Collaboration between parties relevant to the envisioned technology and research conversation is necessary to develop the concept and shape the design format. Knowledge of *technological possibilities and constraints* is essential. So too of the *interaction* that is envisaged between human beings and future technology. There may well be other relevant perspectives, such as legal or ethical viewpoints or domain specialists, that shape the design format too, it depends on the concept and conversation.

- **Culture.** An appreciation of culture, not as in art but in the sense of *shared everyday experience*, is needed to shape the design format. No special expertise is required, social scientists, ethnographers, ethnomethodologists, etc., may be useful but are not necessary. Sensitivity to What Anyone Like Us Knows, i.e., membership competence and our common sense knowledge of the familiar, life as usual scenes that future technology is envisioned to be embedded in and impact, is an essential ingredient in world-building and creating depictions of the future that are familiar and mundane. Most of us possess membership competence, we rely on it to make sense of everyday life too. Leveraging it is essential to crafting depictions of the future that resonate with the here and now.

- **Contiguity.** Crafting design formats so the future is contiguous with the present is crucial. Contiguity legitimates the future, anchoring it in the real world and rendering the real and the fictional "self-consistent" [40]. It is also the key resource for *crafting breaches*. Contiguity may be subtlety manipulated through the design of the experience to create tension in mundane depictions of the future by using technology (real and fictional) to *make familiar scenes strange* and thereby impugn participants common sense understanding of those scenes and how they are ordinarily expected to work. We made the familiar scene of watching TV strange, and thereby impugned participants understanding of how watching TV works, by surfacing the underlying infrastructure (real and fictional) of smart technologies.

There is no requirement that future worlds are designed and breached in the same way we did it, by surfacing normally invisible features of existing or future technologies, or for the same interdisciplinary mix to be employed in shaping those worlds, but sensitivity to **The 4 Cs** – conceptual conversation, collaboration, culture, and contiguity – should prove useful. Beyond that, there is no recipe, no magic formula, no silver bullet for designing-in the breach beyond the requirement that design fictions both depict future and emerging technologies *in* mundane action and, in doing so, are configured to impugn (call into question) common sense knowledge of how the mundane action depicted *works* and is *expected to work*. Making trouble is not necessary, making strange while preserving the familiarity of everyday scenes is. There is, of course, no recipe for doing design fiction either. It is a creative practice, one that lends itself well to crafting breaching experiments but not one that has traditionally been preoccupied with them. Configuring design fiction as breaching experiment is not, then, business as usual under another name, a new "framing" of design fiction [43]. It is a new and original inter- or multi-disciplinary *configuration* of design fiction and contribution to ongoing efforts in HCI to identify acceptability and adoption challenges that confront future and emerging technologies. If others are to adopt the approach it is important to take the original socio-logic of the breaching experiment on board and employ *The 4 Cs* alongside a world-building approach towards experiencing the future mundane.

## 6. Conclusion

We introduced this paper with a quote from Reeves et al. [64] concerning the fragility of technological futures: "even the most mundane of acts can unravel if expected outcomes are not met." The point of the quote was to highlight that the acceptability and adoption of future and emerging technologies turns on their ability to be woven into the "most mundane of acts". If they can't then they "unravel", are undone and ultimately fail as they have no traction in everyday life. The quote also highlights that the success or failure of future and emerging technologies turns, *within* the most mundane of acts, on their ability to meet "expected outcomes", which reflects the common sense and taken for granted understanding that mundane action is expected to bring about particular results. That it is taken for granted is deeply problematic, as it means that expectations about how mundane action should play out are tacit, they sit in the background, unspoken, unremarkable. If we are to identify taken for granted background expectancies that are fateful for the acceptability and adoption of new technology, we need approaches that are capable of surfacing them.

The suggestion here is that configuring design fiction as a breaching experiment may do the job. Breaching experiments are familiar to HCI, used in many and varied ways to 'make trouble' in social situations in a bid to surface issues of relevance to design. We offer an alternative reading here that considers their *sociological function* as originally conceived in the social sciences to open up new methodological opportunities for HCI. Seen from this perspective, breaching experiments were devised as a means of rendering the life as usual character of everyday scenes *strange* for a society's members in order to bring the taken for granted background expectancies they hold and share in common into view. Making the familiar strange for members, not making trouble for them, is key. This does not mean that we only need to make familiar scenes look different, by introducing some fantastic new technology into mundane action for example, to breach what anyone knows. It is essential too that familiar scenes are *designed to impugn* common sense knowledge and thus call what anyone knows about how familiar scenes ordinarily work and are expected work into question. This is not a general feature of design fiction to date.

We have provided a demonstration of our approach, which leverages design fiction as world-building to create an experiential future – the EFM – that enables potential users to get their hands-on and interact with smart adaptive media. The EFM is the first of a kind. It demonstrates that design fiction *can* be configured as a breaching experiment and in doing so underscores an original *methodological* contribution to HCI and efforts to apprehend acceptability and adoption challenges that confront future technology in everyday life. Others may adopt this novel approach by applying *The 4 Cs:* design the breach around the *conceptual conversation* you wish to engage participants in; develop the conceptual conversation through inter- or multi-disciplinary *collaboration* to understand technological possibilities and constraints, the interaction envisaged between human beings and future technologies, and other viewpoints relevant to the conceptual conversation you wish to have; create depictions of the future that reflect shared *cultural* understandings of the familiar scenes future and emerging technologies are to be embedded in; exploit *contiguity* to create tension in mundane depictions of the future by using technology (real and fictional) to make familiar scenes strange and thereby impugn participants understanding of their real circumstances here and now [26].

**Acknowledgements.** This work was supported by the Engineering and Physical Sciences Research Council [grant number EP/S02767X/1]. The human studies reported in this work were approved by the Faculty of Arts and Social Sciences and Lancaster University Management School (FASSLUMS-2022-0762-RECR-3) and School of Computer Science, University of Nottingham (CS-2022-R17).

**Data availability statement.** The data drawn on in the paper is available at http://doi.org/10.17639/nott.7408